\documentclass[useAMS,usenatbib]{mn2e}
\usepackage{amsmath,psfig,epsfig,graphics}
\usepackage{epstopdf}

\def\aap{A\&A}

\def\apj{ApJ}

\def\apjl{ApJ}
\def\mnras{MNRAS}

\def\lesssim{\mathrel{\hbox{\rlap{\hbox{\lower4pt\hbox{$\sim$}}}\hbox{$<$}}}}
\def\gesssim{\mathrel{\hbox{\rlap{\hbox{\lower4pt\hbox{$\sim$}}}\hbox{$>$}}}}

\def\lesssim{\mathrel{\hbox{\rlap{\hbox{\lower4pt\hbox{$\sim$}}}\hbox{$<$}}}}
\def\gesssim{\mathrel{\hbox{\rlap{\hbox{\lower4pt\hbox{$\sim$}}}\hbox{$>$}}}}

\begin{document}

\author[Morandi, Nagai, \& Cui]
{Andrea Morandi${}^1$\thanks{E-mail: amorandi@purdue.edu}, Daisuke Nagai${}^{2,3,4}$, Wei Cui${}^1$\\
$^{1}$ Department of Physics, Purdue University, West Lafayette, IN 47907, USA\\
$^{2}$ Department of Physics, Yale University, New Haven, CT 06520, USA\\
$^{3}$ Department of Astronomy, Yale University, New Haven, CT 06520, USA\\
$^{4}$ Yale Center for Astronomy \& Astrophysics, Yale University, New Haven, CT 06520, USA
}

\title[Measuring the clumpiness in galaxy clusters]
{Non-parametric method for measuring gas inhomogeneities from X-ray observations of galaxy clusters}
\maketitle

\begin{abstract}
We present a non-parametric method to measure inhomogeneities in the intracluster medium (ICM) from X-ray observations of galaxy clusters. Analysing mock \emph{Chandra} X-ray observations of simulated clusters, we show that our new method enables the accurate recovery of the 3D gas density and gas clumping factor profiles out to large radii of galaxy clusters. We then apply this method to \emph{Chandra} X-ray observations of Abell~1835 and present the first determination of the gas clumping factor from the X-ray cluster data. We find that the gas clumping factor in Abell~1835 increases with radius and reaches $\sim 2-3$ at $r=R_{200}$. This is in good agreement with the predictions of hydrodynamical simulations, but it is significantly below the values inferred from recent \emph{Suzaku} observations.  We further show that the radially increasing gas clumping factor causes flattening of the derived entropy profile of the ICM and affects physical interpretation of the cluster gas structure, especially at the large cluster-centric radii. Our new technique should be useful for improving our understanding of the cluster structure and to advance the use of galaxy clusters as cosmological probes, by helping to exploit rich data sets provided by \emph{Chandra} and \emph{XMM}-\emph{Newton} X-ray space telescopes.
\end{abstract}

\begin{keywords}
cosmology: observations -- galaxies: clusters: general -- X-rays: galaxies: clusters
\end{keywords}

\section{Introduction}\label{intro}

Clusters of galaxies are the largest gravitationally bound structures in the universe, promising to serve as powerful laboratories for cosmology and astrophysics. In the current hierarchical structure formation paradigm, galaxy clusters form out of the gravitational collapse of high-density peaks in the primordial density fluctuations in the early universe and grow through mass accretion. Since their evolution traces the growth of linear density perturbations, the changes in the number density of galaxy clusters as a function of time is highly sensitive to the nature of dark matter (DM) and dark energy \citep[e.g.,][]{allen2008,vikhlinin2009,mantz2010}. 

The outskirts of galaxy clusters have become an especially important territory for studying the cosmological growth of structures. They play a key role in cluster cosmology, because these regions are believed to be much less susceptible to astrophysical processes, such as radiative gas cooling, star formation, and energy injection from stars and active galactic nuclei, than the cluster interior. As such, the cluster gas in the outer volume should be tractable with modern N-body+hydrodynamical simulations, since they are governed primarily by hydrodynamics of gas coupled with the collisionless dynamics of DM. 

Recently, \emph{Suzaku} observations have extended X-ray measurements of the intracluster medium (ICM) profiles out to and beyond $R_{200}$, where $R_{200}$ is the radius within which the mean total density is 200 times the critical density of the Universe. Initial results were surprising, revealing significant departures from the theoretical predictions. These observations showed that, at large cluster-centric radii ($r\gesssim R_{200}$), the observed entropy profiles becomes flatter \citep[e.g.,][]{george2009}, and the observed gas fraction exceeded the cosmic baryon fraction of the universe \citep{simionescu2011}. However, these initial results were called into question, because of the difficulty in modelling out point sources and/or galactic foreground with the limited angular resolution ($\gesssim 1$ arcmin) of \emph{Suzaku} at the level required for measuring the extremely low surface brightness in the cluster outskirts. Moreover, the \emph{Suzaku} point-spread function can cause significant contamination from the brightness in the central regions to large radii. Independent \emph{ROSAT} PSPC observations (with low background and large field of view) have shown the steepening of the gas density profile at large radii, but at the level considerably smaller than those inferred from the \emph{Suzaku} observations \citep{eckert2012}. Observational constraints on the gas clumping factor inferred from the joint analysis of the \emph{ROSAT} X-ray and \emph{Planck} Sunyaev-Zeldovich (SZ) observations are also significantly smaller than those inferred from the \emph{Suzaku} measurements \citep{eckert2013}.

In the hierarchical structure formation model, the outskirts of galaxy clusters are virialization regions where galaxies and groups of galaxies accrete, and gaseous component associated with these infalling clumps are stripped and mixed with the surrounding ICM through ram-pressure and tidal stripping, generating the clumpy and turbulent ICM. If not taken into account, these non-equilibrium effects could be a source of significant systematic uncertainties in the X-ray measurements of the ICM profiles as well as global cluster properties. For example, inhomogeneities in the gas distribution lead to the overestimate of the observed gas density and flattening of the entropy profile \citep {nagai2011,zhuravleva2013,roncarelli2013}, which in turn introduces biases in global cluster properties, such as the gas mass fraction \citep{battaglia2012} and the low-scatter X-ray mass proxy, such as $Y_X \equiv M_{\rm gas} T_X$ \citep{khedekar2013}. If they are not properly understood and modelled, these non-equilibrium processes can limit the use of galaxy clusters as cosmological probes. 

At $r=R_{200}\approx 1.6R_{500}$, hydrodynamical simulations predict a gas clumping factor (see Equation~\ref{sx1} in \S~\ref{tecn}) in the range $\sim 1.3-2$ \citep{nagai2011,zhuravleva2013,battaglia2012}. Gas clumping factor inferred from these simulations is in reasonable agreement with the \emph{ROSAT}+\emph{Planck} observations (within the current statistical errors), but they are significantly smaller than those derived from \emph{Suzaku}. Clearly, further work is required to understand the remaining tension in various X-ray measurements and to improve the present observational constraints on gas clumping factor and their implications for the cluster cosmology.

\emph{Chandra} X-ray telescope provides useful insights on this issue. In particular, the \emph{Chandra}'s exquisite ($0.5$ arcsec) angular resolution enables effective identification and removal of point sources, critical for robust measurements of the extremely low surface brightness level in the cluster outskirts, which is well below the cosmic X-ray background. The superb angular resolution of \emph{Chandra} is also essential for detecting small-scale clumps and distinguishing the X-ray emissions arising from clumps and diffuse components. Ultimately, ultra-deep \emph{Chandra} observations of a nearby cluster (e.g., 2.4 Ms Chandra XVP observations of A133 by Vikhlinin et al., in preparation) are what is required to effectively measure X-ray emissions from point sources, gas clump and diffuse ICM gas individually, while minimizing the contribution from the bright cluster core to the emission in the outer volumes. However, such observations are extremely expensive and limited to a very small number of systems in practice. Therefore, in order to investigate the gas clumping phenomena in a larger population of clusters and in a broader redshift range, it is necessary to develop robust statistical techniques to exploit rich data sets available from the \emph{Chandra} and/or \emph{XMM}-\emph{Newton} archives. 

The primary goal of this work is therefore to develop a robust non-parametric approach in order to derive the gas clumping factor from X-ray observations of galaxy clusters. To assess its performance, we analyse mock \emph{Chandra} observations of simulated galaxy clusters extracted from hydrodynamical simulations, taking into account instrumental responses as well as noise due to the Poisson statistics and X-ray background. We then apply our method to \emph{Chandra} observations of Abell~1835 (A1835), a luminous cool-core galaxy cluster at $z=0.253$. A1835 is an optimal cluster for this work as it is one of the most relaxed clusters with deep ($200$~ks) \emph{Chandra} exposure. Moreover, the field-of-view of the ACIS--I chips encompasses $R_{200}$ of this cluster.

The paper is organized as follows. In \S~\ref{tecn} we outline the method. We test the method using mock \emph{Chandra} observations of simulated clusters in \S~\ref{dmicm2} and apply it to real \emph{Chandra} observations of A1835 in \S~\ref{dataan}. Throughout this work we assume the flat $\Lambda$CDM model, with matter density parameter $\Omega_{\rm m}=0.3$, cosmological constant density parameter $\Omega_\Lambda=0.7$, and Hubble constant $H_{0}=100h \,{\rm km\; s^{-1}\; Mpc^{-1}}$ where $h=0.7$. Unless otherwise stated, we report the errors at the 68.3\% confidence level.

\section{Non-parametric method for deprojecting X-ray data}\label{tecn}

X-ray photons are emitted from the hot ICM primarily through the scattering of electrons off of ions via thermal bremsstrahlung process. The X-ray surface brightness is then given by:
\begin{equation}
S_X = \frac{1}{4 \pi (1+z)^4}  \int n_{\rm e} n_{\rm p} \Lambda(T,Z) \, dl\;\;,
\label{1.em.x.eq22}
\end{equation}
where $\Lambda(T,Z)$ is the cooling function, $T$ and $Z$ are the three-dimensional (3D) gas temperature and metallicity, $n_{\rm e}$ is the electron density and $l$ is the line of sight. Since $\Lambda(T,Z)$ depends weekly on $T$ and $Z \approx 0.3$ for hot ($T_X\gesssim 3$~keV) clusters, the observed X-ray surface brightness profile depends primarily on the average of the gas density squared, 
\begin{equation}
S_X(r) \propto \langle \rho_{\rm gas}^2(r) \rangle = C(r) \langle \rho_{\rm gas}(r) \rangle^2,
\label{sx1}
\end{equation}
where the $C$ is the clumping factor given by 
\begin{equation}
C\equiv\frac{\left < n_{\rm e} ^2\right > }{\,\left < n_{\rm e} \right >^2} \ge 1.
\label{clump}
\end{equation}
Note that $C=1$ if the ICM is not clumpy (i.e. a single phase medium characterized by a single temperature and gas density within each radial bin). In the X-ray cluster analyses, it is commonly assumed that $C=1$, and the three-dimensional gas density distribution is derived from the observed X-ray surface brightness profile by inverting Equation~\ref{sx1}. However, if the ICM is clumpy, the gas density inferred from the X-ray surface brightness is overestimated by $\sqrt{C(r)}$ and the gas entropy $S \equiv T/n_e^{2/3}$ is underestimated by $C(r)^{1/3}$. Note further that a radially dependent clumping factor causes flattening of the derived gas density and entropy profiles, if it is not properly taken into account. 

Given that we only observe the two-dimensional (2D) projected surface brightness in the plane of the sky, it is generally non-trivial to derive the intrinsic 3D physical properties, such as the gas density and gas clumping factor $C(r)$ from clusters observations. In the following section, we present a novel method to measure the 3D gas density and gas clumping factor profiles from X-ray observations of galaxy clusters.

\subsection{Deprojecting 3D gas density}\label{apecdepte}

\begin{figure}
\begin{center}
\psfig{figure=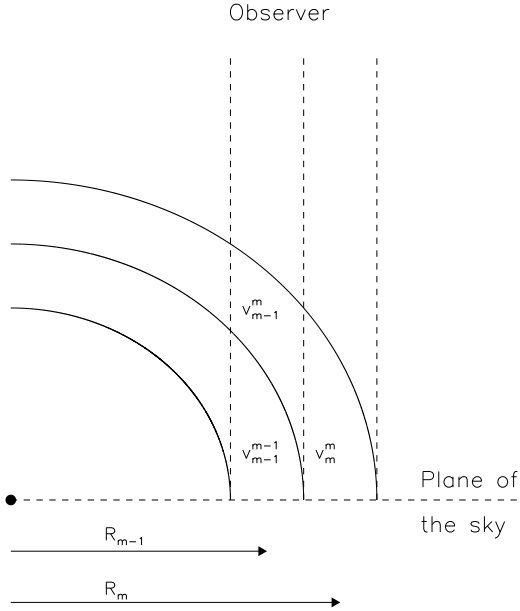,width=0.48\textwidth}
\caption{Illustration of the onion-skin model adopted for the geometrical deprojection. We assume that the cluster is spherically symmetric and it has an onion-like structure, with $n$ concentric spherical shells and $n$ rings or annuli. The matrix ${\bf \sf V_i^j}$ is an upper triangular matrix $n\times n$ matrix, whose entries represent the volume of the $j$-th shell contained inside the $i$-th annulus (with $j \ge i$). The X-ray boundary (the most external annulus) is the $n$-th ring with the area $A_n$ and radius $R_n$.}
\label{fig1}
\end{center}
\end{figure}

The first step in the X-ray cluster profile analysis is to recover the electron density $n_e(r)$ by deprojecting the X-ray surface brightness profile. We assume that the cluster is spherically symmetric, and it has an onion-like structure with $n$ concentric spherical shells, each characterized by uniform gas density and temperature within it (see Fig. \ref{fig1}). Therefore, the cluster image in projection is divided into $n$ rings (or annuli) of area ${\bf A}=(A_1,A_2,...,A_n)$, which are assumed to have the same radii of the 3D spherical shells of radius ${\bf r}=R_i,\; i=1,...,n$. Hereafter we assume that the index $j$ ($i$) indicates the shell (ring) defined by two radii $(r_{\rm in},r_{\rm out})$. The X-ray boundary (the most external ring) is the $n$-th ring with the area $A_n$ and radius $R_n$.
 
From the two-dimensional surface brightness map ${\bf \sf S_X}$, we first compute the 1D surface brightness profile $S_X(r)$ (i.e. the surface brightness averaged in circular annuli). We then converted the surface brightness into the summation of volume emission density under the assumption of spherical symmetry and onion--like structure. Specifically, following \cite{1983ApJ...272..439K}, we express the relation between the surface brightness $S_X$ and the gas density $n_e$ in Equation~\ref{1.em.x.eq22} using the following matrix formalism:
\begin{equation}
S_{X,i}=\frac{1}{4 \pi (1+z)^4} \Lambda(T^*_{\rm proj},Z) \; \sum_{j=i}^n {\bf \sf V}_i^j\left({C_j n_{e,j}^2}\right)/{\bf A}_i+{\bf \sf N}_i,
\label{kk}
\end{equation}
where $\Lambda(T^*_{\rm proj},Z)$ is the cooling function, $T^*_{\rm proj}$ is the observed (projected) temperature, $Z$ is the metallicity, $C=C(r)$ is the gas clumping factor, and $n_e=<n_e>$ is the average value of the gas density in the $j$-th shell.
${\bf \sf N}_i$ is the noise vector at the $i$-th annulus (given by Equation~\ref{eqn:mm4dd}), and ${\bf \sf V}_i^j$ is an element of the upper triangular $n\times n$ matrix with the $n$-dimensional column vectors ${\bf  \sf V^1,V^2, ...,V^n}$ representing the effective volumes, i.e. contributions of the volume fraction of the $j$-th spherical shell to the $i$-th annulus with $j\le i$. Note that we did not use a geometrical volume to deproject the surface brightness, since this assumes a constant gas density in each shell, which introduces a bias in the derived 3D quantities. Instead, we used the effective volume ${\bf \sf V}$, which takes into account the gradient of the physical parameters within each shell in order to provide an unbiased estimate of the 3D gas density \citep[see Appendix~B of][]{morandi2007a}.

Next, we derive the emission measure $\tilde{K}= \int n^2_{\rm e}\, dV$ by solving Equation~\ref{kk} and using a MEKAL model \citep{1992Kaastra, 1995ApJ...438L.115L} for computing the emissivity $\Lambda(T^*_{\rm proj},Z)$. To obtain the emission measure, we create \emph{Chandra} spectra, in which the emissivity model is folded through the instrumental responses (ARF and RMF) along with absorption, temperature and metallicity measured in the $i$-th ring. The model of $S_X$ is then obtained by rescaling the \emph{Chandra} spectra by the observed counts in the $i$-th ring. Finally, we invert the emission measure profile to derive the desired 3D gas density profile. 

Note that this method exploits the mild dependence of the cooling function $\Lambda(T)$ on $T$\footnote{The integrated cooling function in $(0.5-5)$~keV band is approximately given by $\Lambda(T) \propto T^{-\alpha}$ with $0.1\la \alpha\la 0.2$ for $T\sim 7- 12$ keV.}, such that the systematic uncertainties in the estimated projected temperature do not introduce large systematic errors (less than a few percent) in the determination of the normalization of the emission measure at the $i$-th radial shell, $K_i$. This approach thus enables us to derive the projected density in annuli even with modest (e.g., several hundred) counts without using spatially-resolved spectral analysis, which requires large photon statistics (e.g., over $2000$ net counts per annulus). Errors are computed by performing 1000 Monte Carlo simulations of the observed counts. 

\subsection{Constraining the gas clumping factor}\label{apecdepte4}

\begin{figure*}
\begin{center}
 \hbox{
\psfig{figure=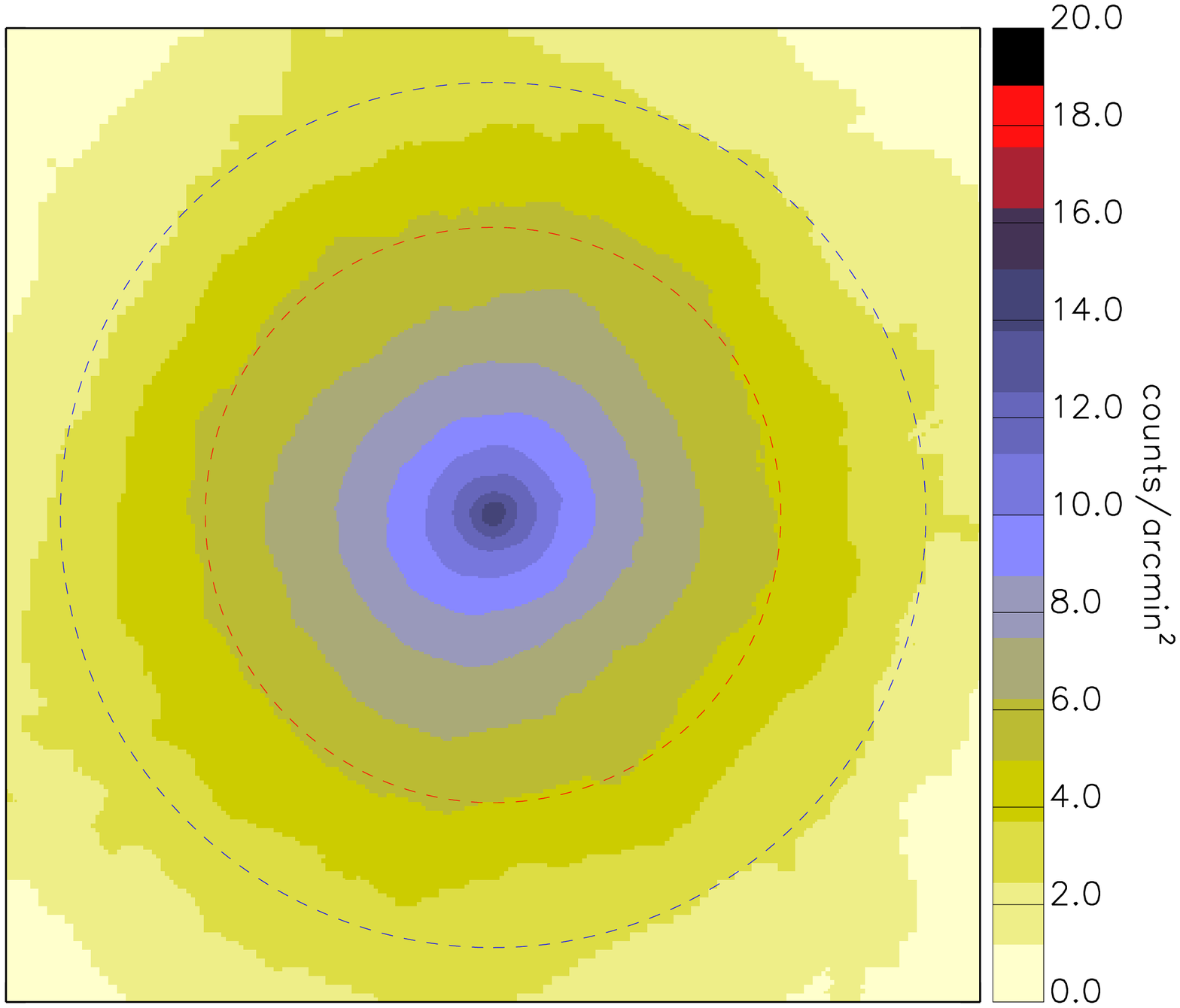,width=0.33\textwidth}
\psfig{figure=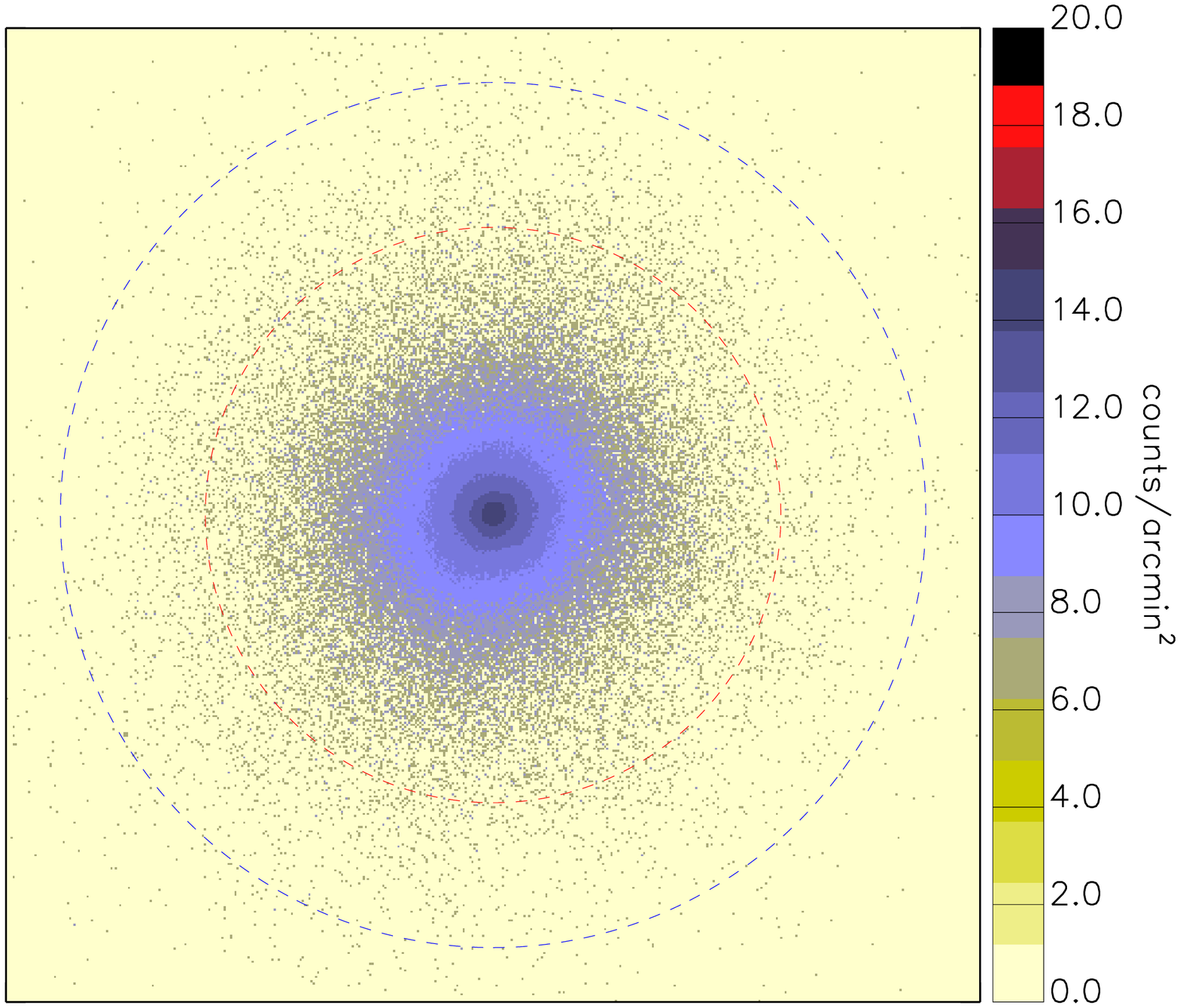,width=0.33\textwidth}
\psfig{figure=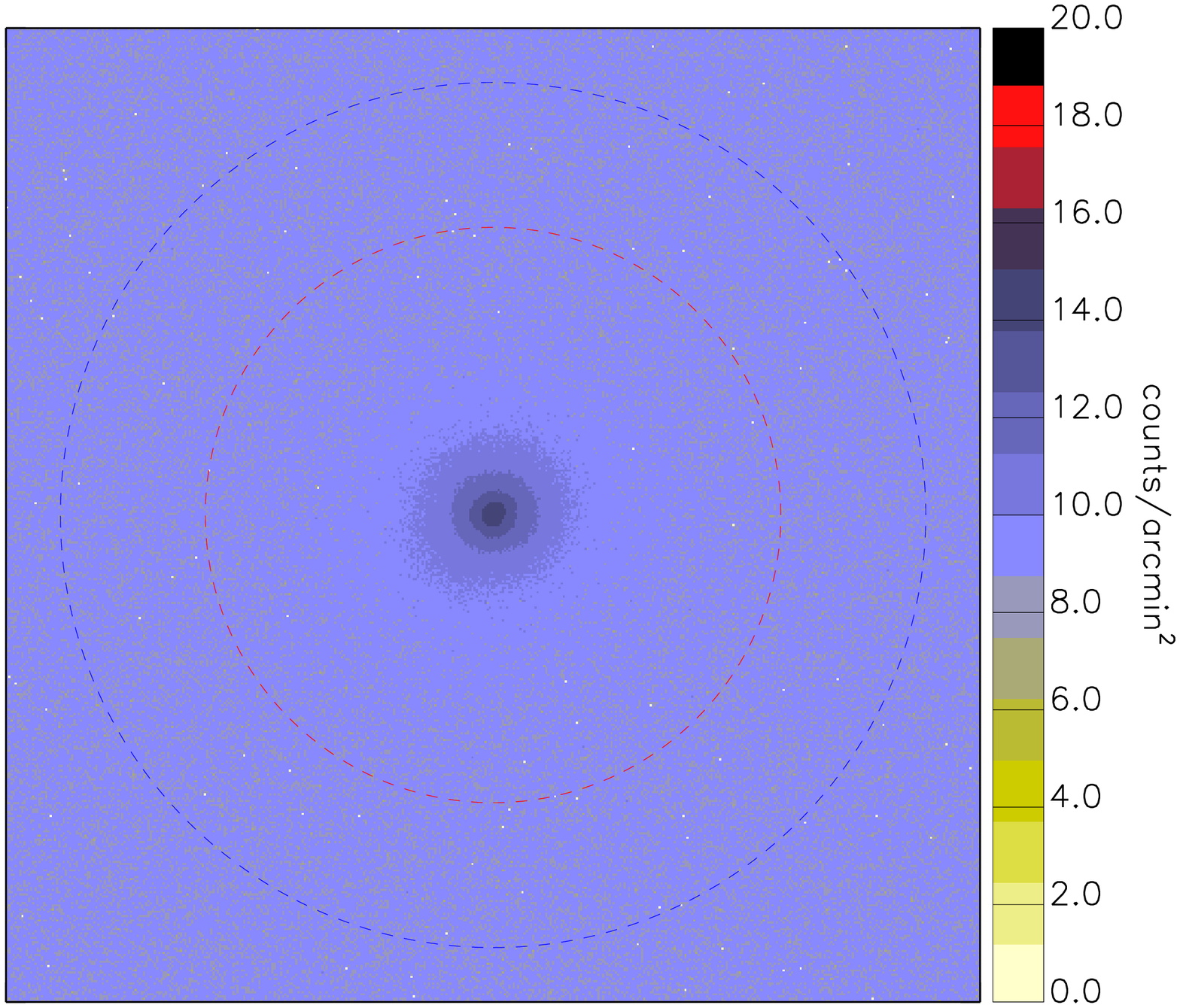,width=0.33\textwidth}
}
\caption[]{Mock \emph{Chandra} simulations of the cluster CL104 viewed along the $z$ projection for a deep ($t_{\rm int}=2\, {\rm Ms}$) integration time. In the left image we show the X-ray emission from the clusters folded through the \emph{Chandra} response without Poisson noise and X-ray background. We then added Poisson noise in the middle image and X-ray background in addition in the right image. The size of the region shown is approximately 4.7~Mpc centred on the minimum of cluster potential. The pixel size of the image is about 10~kpc. Red and blue circles indicate $R_{500}$ and $R_{200}$, respectively. }
\label{entps3xkn3e}
\end{center}
\end{figure*}

Inhomogeneous 3D gas density distribution leaves its imprints on the 2D surface brightness distribution ${\bf \sf S_X}$. From the binned surface brightness map averaged in circular annuli, we can measure the total scatter $\sigma_{S_X,{\rm tot}}$ as a standard deviation of the surface brightness in each annulus:
\begin{equation}
\sigma_{S_X,{\rm tot}}^2=\sigma_{S_X,{\rm intr}}^2+\sigma_{S_X,{\rm noise}}^2 \ ,
\label{eqn:mm4dd}
\end{equation}
which is the sum of the intrinsic scatter in the binned surface brightness ($\sigma_{S_X,{\rm intr}}$) and the noise ($\sigma_{S_X,{\rm noise}}$). The second term includes the contributions from Poisson noise, i.e. the square-root of the total (source+background) counts. The first term contains the information we are looking for; namely, it is related to the scatter in the gas density distribution $\sigma_{n_e,{\rm intr}}$, which in turn is related to the gas clumping factor:
\begin{equation}
C\equiv\frac{\left < n_e ^2\right > }{\,\left < n_{\rm e} \right >^2} = 1+ \frac{\sigma_{n_e,{\rm intr}}^2}{\left<n_e\right >^2} \ ,
\end{equation}
(c.f. Equation~\ref{sx1}). The covariance matrix ${\bf \sf C}_{\rm n_e^2}$ of $n_e^2$ is then related to $\sigma_{S_X,{\rm intr}}$ via the following relation:
\begin{equation}
{\bf \sf C}_{n_e^2} = ({\bf \sf V}^{\rm t}{\bf \sf C}_{S_X}^{-1} {\bf \sf V})^{-1} \ ,
\label{eqn:mm4}
\end{equation}
where ${\bf \sf C}_{S_X}$ is the intrinsic scatter diagonal matrix $\left[\sigma_{S_X,{\rm intr}}\right]^{2}$, with entries on the main diagonal $\sigma_{S_X,{\rm intr},1},\sigma_{S_X,{\rm intr},2},\,...\sigma_{S_X,{\rm intr},n}$ \citep[see e.g.][]{morandi2011b}. Therefore, the idea of our approach is to infer the intrinsic scatter of the gas density distribution $\sigma_{n_e,{\rm intr}}$ from the observed surface brightness inhomogeneities $\sigma_{S_X,{\rm tot}}$ through these relations. 

In practice, the inference of $\sigma_{S_X,{\rm intr}}$ is complicated in presence of noise in real data. Fig. \ref{entps3xkn3e} shows the surface brightness maps ${\bf \sf S_X}$ for the noiseless data (left panel), mock $2\, {\rm Ms}$ \emph{Chandra} observation of a simulated cluster without (middle panel) and with X-ray background (right panel). These figures illustrate that the total (observed) scatter ($\sigma_{S_X,{\rm tot}}$) is the superposition of the intrinsic scatter in the X-ray surface brightness ($\sigma_{S_X,{\rm intr}}$) and noise ($\sigma_{S_X,{\rm noise}}$). The latter becomes larger at the outer radii, where Poisson noise and/or X-ray background are more dominant. 

In order to recover $\sigma_{S_X,{\rm intr}}$, we re-bin the observed surface brightness map ${\bf \sf S_X}$ such that (a) there are enough ($\gesssim15-20$) counts per pixel to assume Gaussian total errors, (b) X-ray background does not show any appreciable spatial variation, and (c) there are enough spatial resolution to identify and remove significant contaminations due to point sources and substructures. Note that the removal of detectable substructures eliminates the non-Gaussian tails of the probability density distributions \citep{zhuravleva2013}, making it approximately Gaussian. We evaluate our ability to significantly measure $\sigma_{S_X,{\rm intr}}$ by using the following {\it F}-test: 
\begin{equation}
F=\frac{\sigma_{S_X,{\rm tot}}^2}{\sigma_{S_X,{\rm noise}}^2} \ ,
\label{eqn:mm4ddb}
\end{equation}
which has the F-distribution under the null hypothesis (i.e. the total scatter is consistent with Poisson noise). In other words, the evidence becomes increasingly inconsistent with the null hypothesis as the value of $F$ becomes larger than 1. In our work, $F$ increases as we increase the annulus size and/or reduce the error variance by taking longer integration time. 

In the following sections we first test our method using hydrodynamical simulations and then apply it to real \emph{Chandra} observations of A1835.

\section{Testing with hydrodynamical simulations}\label{dmicm2}

\subsection{Simulated clusters sample}\label{sim_sample}

We analysed a sample of high-resolution hydrodynamical simulations of galaxy clusters formation from \citet[][hereafter N07]{nagai2007a,nagai2007b}, which were performed using the ART code \citep{kravtsov2002,rudd2008}. In this work, we analysed the outputs of the simulations that include radiative cooling, star formation, metal enrichment and stellar feedback. In particular, we focused on two X-ray luminous clusters, CL104 and CL101, at z=0 with the (core-excised) X-ray temperature of $T_{\rm X}=7.7$ and $8.7$~keV, respectively. CL101 is a massive, dynamically active cluster, which has recently experienced violent mergers (at $z\sim 0.1$ and $z\sim 0.25$) and contains two major substructures near the core at z=0. These two substructures have been identified by visual inspection and masked out before its analysis. CL104 is a similarly massive cluster, but with a more quiescent mass accretion history. This cluster has not experienced a significant merger for the past 8~Gyr, making it one of the most relaxed systems in the N07 sample. Each cluster is simulated using a $128^3$ uniform grid with eight levels of refinement. Clusters are selected from 120$h^{-1}$~Mpc computational boxes, achieving peak spatial resolution of $\approx 3.6h^{-1}$~kpc, sufficient to resolve dense gas clumps on scales larger than the resolution limit. The DM particle mass in the region surrounding the cluster is $9\times 10^8 h^{-1}M_{\odot}$, while the rest of the simulation volume is followed with lower mass and spatial resolution. We refer readers to N07 for further details of the simulations. 

We generate mock $2\, {\rm Ms}$ \emph{Chandra} observations of the simulated clusters, by placing them at the observing redshift of $z=0.253$. We then analysed the outputs of the mock \emph{Chandra} spectra of the simulated clusters. Specifically, from the 3D gas density, temperature and metallicity data cubes, we first compute the X-ray surface brightness map ${\bf \sf S_X}$ using Equation~\ref{1.em.x.eq22}. For each cluster and for each projection axis ($x$, $y$ and $z$), we then create mock \emph{Chandra} spectra, using the MEKAL model for the emissivity $\Lambda(T^*_{\rm proj},Z)$ and convolving it with the instrumental responses (ARF and RMF). The simulated spectroscopic temperatures are obtained by fitting the mock spectra in the $(0.6-7)$~keV range using a single temperature MEKAL model after subtracting the background. We assumed the same hydrogen column density and the ACIS ``blank-sky" background used in the analysis of A1835 discussed in \S~\ref{laoa}. We estimate errors at the 68\% confidence level using the tasks {\it error} and {\it steppar}.

\begin{figure*}
\begin{center}
 \hbox{
\psfig{figure=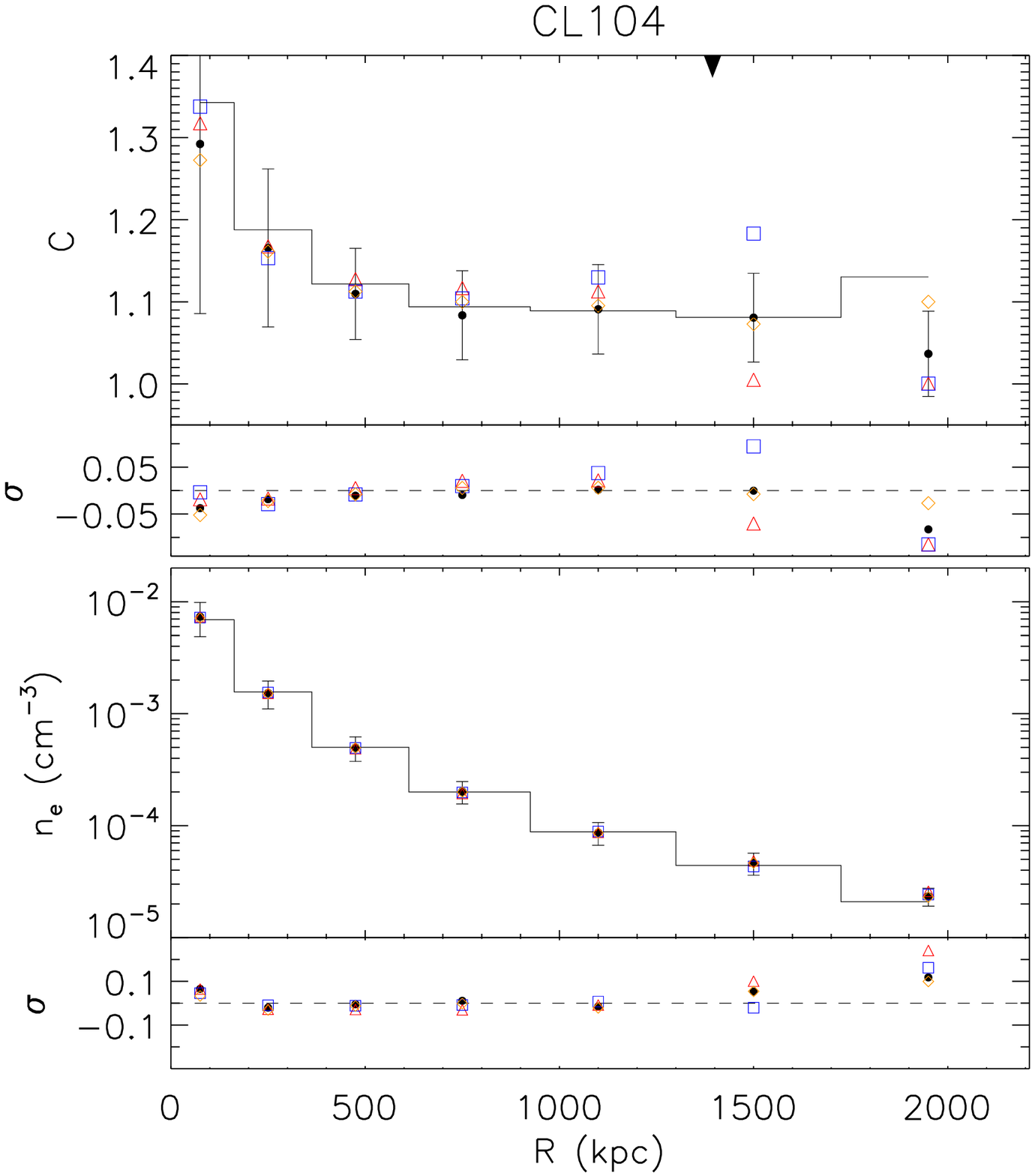,width=0.5\textwidth}
\psfig{figure=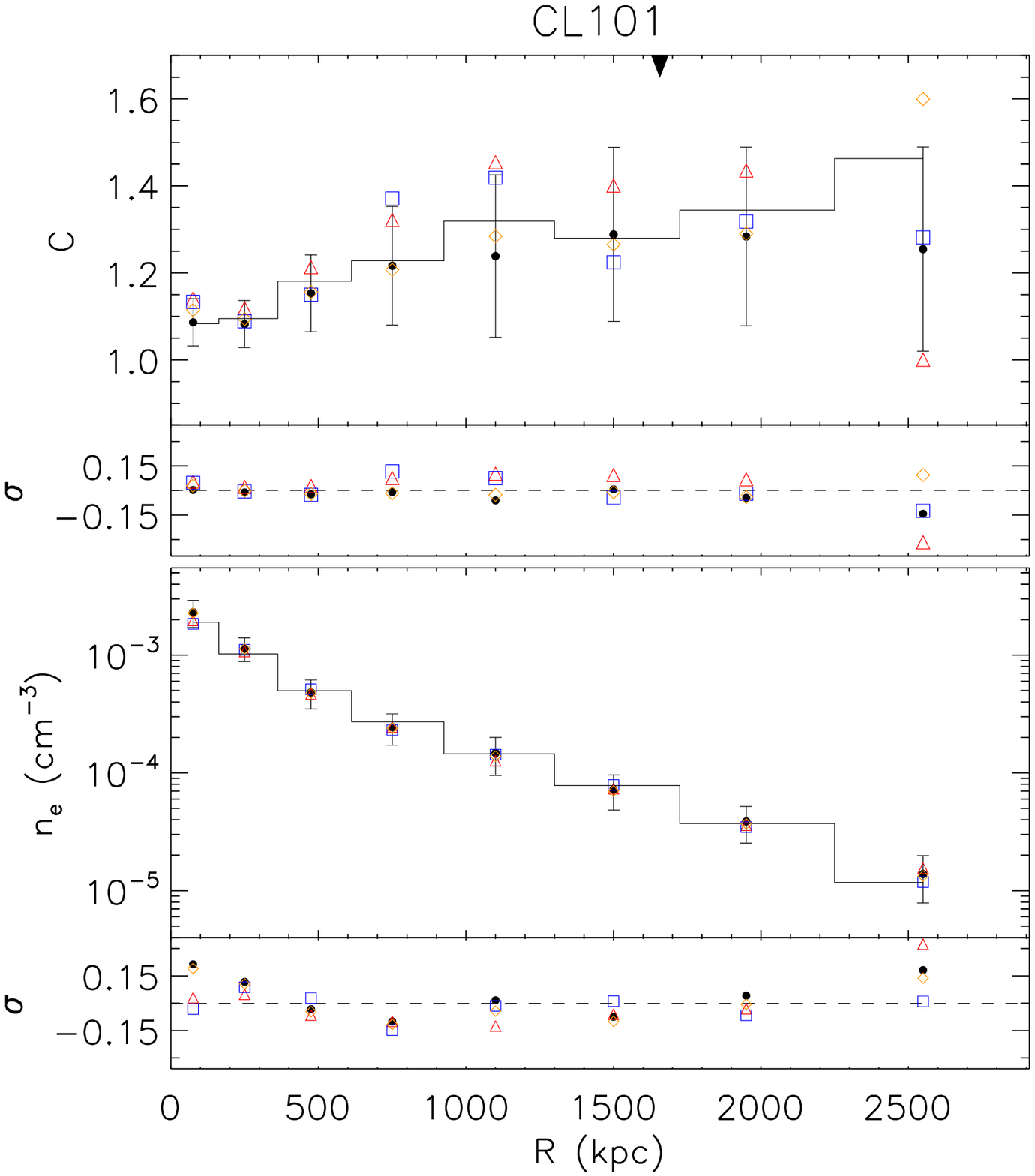,width=0.5\textwidth}
}
\caption[]{Three-dimensional gas clumping (upper panel) and gas density (lower panel) profiles retrieved from mock \emph{Chandra} analyses of CL104 (left panel) and CL101 (right panel). Solid line shows the true clumping factor (in the top panel) and gas density (in the bottom panel) of the clusters in the $i$th spherical shell, whose width is graphically represented by the horizontal solid lines. Circles with errorbars indicate the gas clumping factor (in the top panel) and gas density (in the bottom panel) retrieved from the X-ray analysis along the $z$ projection, while the triangles (squares) indicate the $x$ ($y$) projection (errorbars are omitted for clarity). We also report the fractional total scatter at the bottom of each panel. Diamonds show the results of the noiseless analysis (i.e. neglecting Poisson noise and X-ray background) for the $z$ projection. The arrowhead pointers at top of each plot indicate the locations of $R_{500}$.}
\label{entps3xkn3}
\end{center}
\end{figure*}

\subsection{Reconstruction of gas density and clumpiness from simulations}\label{phys11c}

We test the performance of our non-parametric method for reconstructing the 3D gas clumping factor using mock \emph{Chandra} simulation of galaxy clusters. We re-binned the observed surface brightness map ${\bf \sf S_X}$ so that there are enough counts ($\gesssim15-20$) per pixel to assume Gaussian total errors. We then compute the azimuthally-averaged surface brightness profile $S_X(r)$ and derive the 3D gas density and gas clumping factor profiles using the method described in \S \ref{apecdepte}. We selected the boundary radius according to the following criteria: the null hypothesis that the total scatter is consistent with Poisson noise is rejected with a probability of 90\% (see Equation~\ref{eqn:mm4ddb}).

Fig.~\ref{entps3xkn3} shows the recovery of 3D gas density and clumping factor profiles from the $2\, {\rm Ms}$ mock \emph{Chandra} analysis of CL101 and CL104 clusters viewed along three orthogonal projections (indicated by triangle, square, circle points for $x, y, z$ projection axes, respectively). In addition, we performed a noiseless analysis without Poisson noise and X-ray background (indicated by diamonds). We find a good agreement between the measured gas density and gas clumping factor profiles and their true values measured directly in simulations out to $r \sim 1.5 R_{200}$. For the relaxed cluster CL104, the recovery of the 3D gas density is accurate to $5\%-10\%$. The recovery is still quite good for the unrelaxed cluster CL101, with a slightly enhanced bias at the level of $15\%-20\%$ due to the combination of substructures and aspherical gas distribution in such a disturbed system. 

\section{X-ray analysis of Abell~1835}\label{dataan}

The cluster A1835 is a luminous cluster at redshift $z=0.253$, which exhibits several indications of a relaxed cluster. For instance, its X-ray emission peak is associated with a cool core and it is well centred on the Brightest Cluster Galaxy. The X-ray isophotes appear quite regular, with a low degree of ellipticity, and there are no significant substructures. This is also a strong cooling core (SCC) cluster, where the central cooling time $t_{\rm cool}\simeq 10^9$~yr is considerably less than the age of the universe \citep{morandi2007b}. The cool-core-corrected X-ray temperature is $T_{\rm X}=10.12\pm0.15$~keV, and the abundance is $Z=0.44\pm0.04\, Z_{\odot}$. Like other SCC clusters, A1835 shows a strong spike in the X-ray surface brightness profile and a drop in the temperature with $T\lesssim 5$~keV in the central region. The temperature profile is also quite regular throughout the cluster. A full description of the X-ray analysis can be found in \cite{morandi2012a}. Here we briefly summarize aspects of our data reduction and analysis of A1835 relevant for this work.

\subsection{X-ray data reduction}\label{laoa}
We reduced the \emph{Chandra} X-ray data using the \textit{CIAO} data analysis package -- version 4.5 -- and the calibration data base CALDB 4.5.6. We analysed three data sets retrieved from the NASA HEASARC archive (observation ID 6880, 6881 and 737 obtained by using the ACIS--I CCD imaging spectrometer) with the total exposure time of approximately 200~ks. 

The level-1 event files were reprocessed to apply the appropriate gain maps and calibrations. We used the {\texttt{acis\_process\_events}} tool to check for cosmic ray background events and to correct for eventual spatial gain variations caused by charge transfer inefficiency to re-compute the event grades. We then filtered the data to include the standard event grades 0, 2, 3, 4 and 6 only, and thereby filtering for the Good Time Intervals (GTIs). We then used the tool {\tt dmextract} to create the light curve of the local background, followed by a careful screening to discard contaminating flare events. In order to clean the data sets associated with the periods of anomalous background rates, we used the {\tt deflare} script, filtering out the times where the background count rate exceeds $\pm 3\sigma$ of the mean value. Finally, we select the energy range $0.3-12$~keV to generate a level-2 event file.

Bright point sources ($\sim 100$) were identified and masked out using the script \texttt{vtpdetect} from the level-2 file. In addition, we masked out two additional low-mass clusters identified by the Sloan Digital Sky Survey (SDSS), MAXBCG J210.31728+02.75364 and WHL J140031.8+025443 \citep[see][]{bonamente2013}. After masking out point sources and low-mass systems, the X-ray images were extracted from the level-2 event files in the energy range ($0.5-5.0$)~keV, corrected for the exposure map to remove the vignetting effects. We then create an exposure-corrected image from a set of observations using the \texttt{merge\_obs} to combine three ACIS--I observations. All maps were checked by visual inspection at each stage of the process.

From the X-ray images we measure the emission measure profile, which is then inverted to obtain the 3D gas density profile. We first determine the centroid ($x_{\rm c},y_{\rm c}$) of the surface brightness image by locating the position where the derivatives of the surface brightness variation along two orthogonal (e.g., X and Y) directions become zero, which was found to be a more robust determination of the centroid than the centre of mass or fitting a 2D Gaussian. We constructed a set of circular annuli around the centroid of the surface brightness. In each concentric ring, we performed a spectral analysis (a) by extracting the source and background spectra in the energy range $(0.6-7)$~keV ($(0.6-5)$~keV in the last annulus) using the \emph{CIAO} \texttt{specextract} tool from each observation and (b) by fitting the composite spectra using the APEC emission model \citep{foster2012}. For the spectral fitting, we group photons into bins of at least 20 counts per energy channel and applying the $\chi^2$-statistics. Thus, there are three free parameters: the normalization of the thermal spectrum $K_{\rm i} \propto \int n^2_{\rm e}\, dV$, the emission-weighted temperature $T^*_{\rm proj,i}$, and the metallicity $Z_{\rm i}$. We assume the HI column density of $N_H=0.0232\times 10^{22} {\rm cm}^{-2}$. We analysed three observations individually to check for consistency before analysing a joint data set. 

Since we are analysing ACIS imaging observations whose field-of-view is comparable to the extent of the source, it is difficult to estimate the background from the same data set. We therefore estimate the background spectra using the ACIS ``blank-sky" background files as follows. We first extract the blank-sky spectra from the blank-field background data provided by the ACIS calibration team in the same chip regions as in the observed cluster spectra. We then analyse the blank-sky observations using the data reduction procedure consistent with the one applied to the cluster data. Specifically, we re-project the blank-field background data on to the sky according to the observation aspect information using the {\tt reproject\_events} tool. We then scale the blank-sky spectrum level to the corresponding observed spectrum in the 9.5-12 keV energy range, in which the contribution from the cluster emission is small. In this way, we are using the ARF and RMF files derived consistently for the source and the estimated background. However, in the soft X-ray band ($\lesssim 2$~keV), appropriate adjustments are required since the background can vary both in time and in space \citep{vikhlinin2006}. To make these adjustments, we extract spectra from a certain fraction of the detector areas that are free from the cluster emission and subtract the adjusted blank-field background. We then fit the residuals with an unabsorbed $\sim 0.5$ keV thermal plasma model at $z=0$ and assuming solar metallicity. We include the recovered soft emission as an additional component in the spectral fits. 

\begin{figure*}
\begin{center}
 \hbox{
\psfig{figure=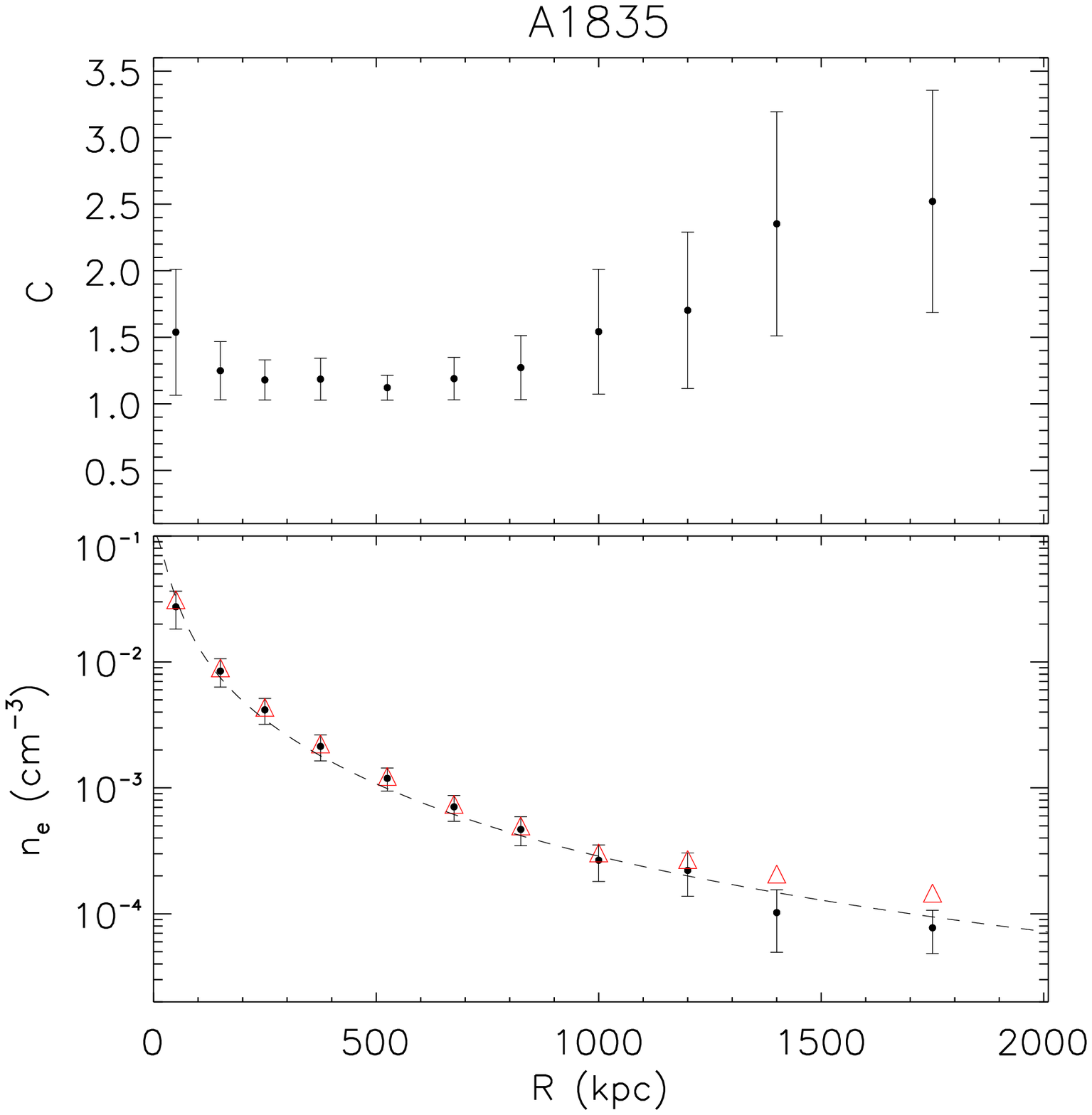,width=0.5\textwidth}
\psfig{figure=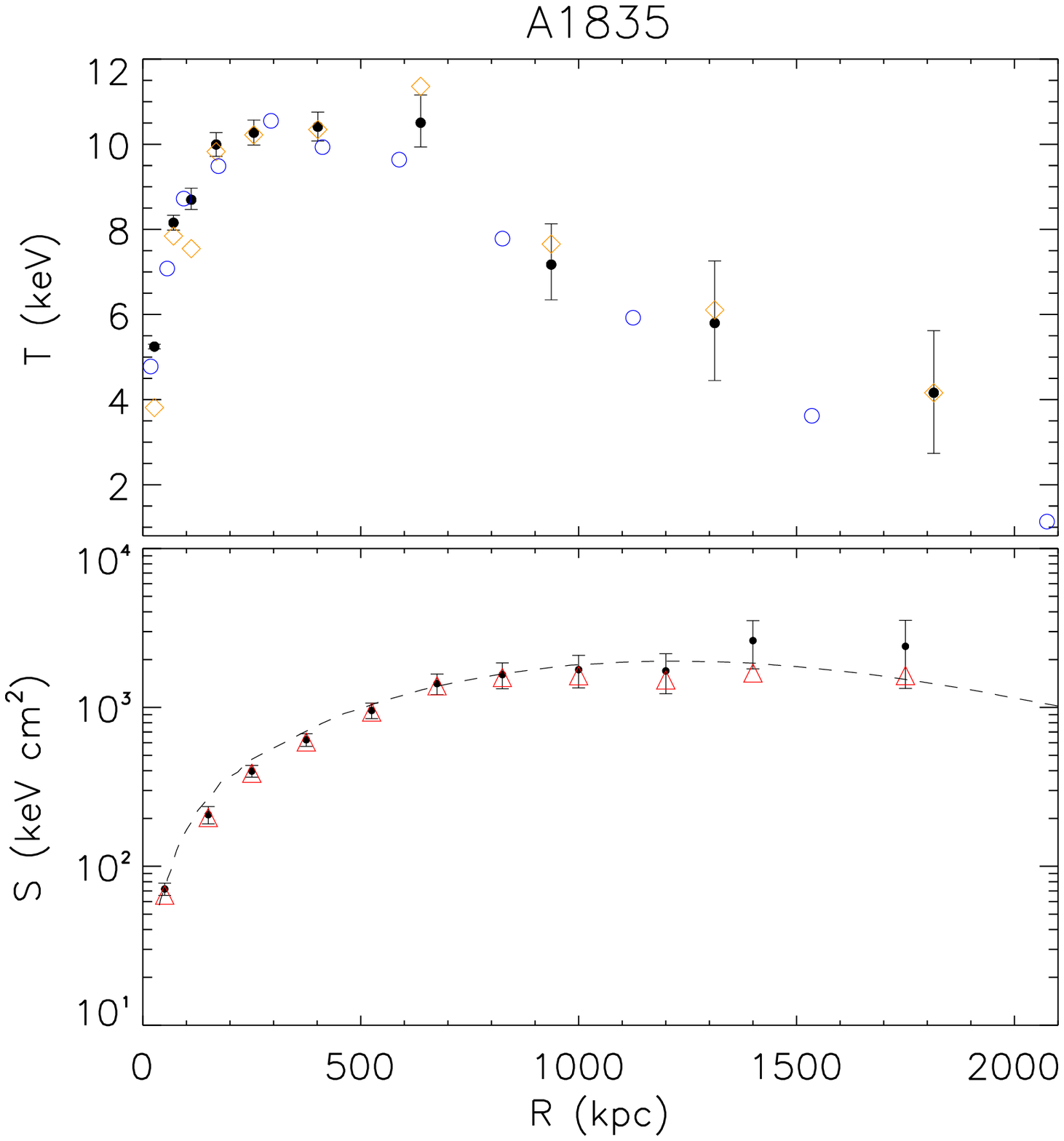,width=0.5\textwidth}
}
\caption[]{\emph{Chandra} X-ray observations of A1835. {\it Left panel:} 3D gas clumping factor (upper panel) and gas density (lower panel) profiles. Dashed line shows the 3D gas density profile from \citet{bonamente2013}. {\it Right panel:} observed and deprojected temperature profile (upper panel) and entropy profile (lower panel). In the bottom panels, we show results with (circles) and without (triangles) the gas clumping factor, while the dashed line indicates the profiles from \citet{bonamente2013}. In the top-right panel, the solid circles and diamonds show the projected and deprojected temperature profiles, respectively, while the blue open circles are the projected temperature profile from \citet{bonamente2013}. For the deprojected temperature we omitted the errorbars for clarity. $R_{500}=1587$~kpc.
}
\label{entps3xkn3fr}
\end{center}
\end{figure*}

\subsection{Reconstruction of gas clumpiness and density in A1835}\label{phys118d}
We re-bin the stacked surface brightness map ${\bf \sf S_X}$ into cells of $16\times16$ pixels each, such that there are more than $15$ counts per pixel. We emphasize that we rebin the surface brightness image after removing point sources from the original \emph{Chandra} map with about $0.5$~arcsec angular resolution. This enables effective identification and removal of point sources, which can significantly affect the extremely low surface brightness level in the cluster outskirts, which is well below the cosmic X-ray background. We then compute the azimuthally averaged surface brightness profile $S_X(r)$. Finally, we apply the method described in \S \ref{apecdepte} on $S_X(r)$ to derive the 3D gas density and gas clumping factor profiles. We selected the boundary radius according to the following criteria: the null hypothesis that the total scatter is consistent with Poisson noise is rejected with a probability of 90\% (see Equation~\ref{eqn:mm4ddb}). The boundary radius is set to 2100 kpc in our A1835 analysis.

The left panel of Fig. \ref{entps3xkn3fr} shows the derived 3D clumping factor and gas density for A1835. The gas clumping factor profile becomes larger than the unity at $r \gesssim R_{500}$, where $R_{500}=1587$~kpc. The gas clumping factor appears to increase with radius, reaching $C \approx 2-3$ at $r\gesssim R_{500}$, but the error bar is large. The right panel of Fig. \ref{entps3xkn3fr} shows the temperature profile of A1835, which shows a drop by roughly a factor of 2.5 from the peak temperature at $r\approx R_{200}$. Our finding is analogous to the recent result by \citet{bonamente2013} who also reported a drop in temperature profile at large radii.  However, we find that our temperature drop is less pronounced than their measurements.  As a result, our derived entropy profile is nearly flat at large radii ($r\gesssim R_{500}$), in contrast to the recent claim by \cite{bonamente2013} who reported a negative entropy gradient beyond $R_{500}$, which led the authors to conclude that the cluster gas is convectively unstable at large radii. We note that this is due to the combination of differences in the derived projected temperature profile as well as the gas clumping factor considered in this work, but not in \cite{bonamente2013}. Further work is needed to understand the origin of these discrepancies. Nevertheless, we verified that these systematic uncertainties have a small impact on the recovery of the 3D gas density and hence gas clumping factor profiles, as evidenced by the good agreement between our gas density profile (without gas clumping factor correction) and that derived by \cite{bonamente2013} (see the lower-left panel of Fig. \ref{entps3xkn3fr}).

We point out that the angular resolution of the rebinned image ($\approx 8$~arcsec, which corresponds to the physical scale of about $30$~kpc) sets the smallest scale down to which we can resolve gas clumps.  Note also that, in \S~\ref{phys11c}, we also tested our method on the simulated clusters which resolves dense gas clumps down to about $10$~kpc, demonstrating that we can indeed recover the true clumping factor in the simulations from the rebinned image (see Fig.~\ref{entps3xkn3}). However, since our method does not detect the small-scale clumps below $10$~kpc, we emphasize that the clumping factor inferred from our method should be taken as a {\it lower limit} of the true clumping factor.



\section{Discussions \& Conclusions}\label{conclusion33}

In this work we present a non-parametric method for reconstructing the gas clumping factor in the ICM from X-ray observations of galaxy clusters.  For the first time, our method makes a direct and logical connection between the X-ray surface brightness fluctuations and the 3D clumping factor.  Analyzing mock \emph{Chandra} observations of simulated clusters, we show that this method enables the accurate recovery of the 3D gas density and clumping factor profiles out to large radii of galaxy clusters. We then use this method to derive a first observational measurement of the gas clumping factor out to $R_{200}$ from \emph{Chandra} X-ray observations of A1835. The \emph{Chandra}'s superior angular resolution enables robust identification and removal of point sources from the X-ray images, while minimizing the contribution from the bright cluster core to the emission in the outer volumes. We find that the gas clumping factor increases with radius and reaches $2-3$ at $r\gesssim R_{500}$ of A1835. Our results are in good agreement with the predictions of hydrodynamical simulations, but significantly smaller than those inferred from \emph{Suzaku}. 

We also show that the radially increasing gas clumping factor also causes flattening of the entropy profile and affects physical interpretation of the stability of the cluster gas structure at large radii. However, we find that the observed entropy profiles in the outskirts of A1835 cannot be explained by the observed clumpiness alone. Our analysis shows that the observed temperature profile decreases monotonically with radius towards the outskirts of A1835, and this is also partly responsible for flattening of the observed entropy profile at $r\gesssim R_{500}$.  Additional cluster physics, such as non-thermal pressure due to bulk and turbulent gas motions, is likely required to explain these observed temperature and entropy profiles \citep{fuscofemiano2013}. Note, however, that our results are marginally in tension with previous findings, and further work (with special care on the treatment of foreground and background) is needed to evaluate the relative importance of gas clumping and non-thermal pressure support from these data. 

\section*{acknowledgements}
We thank Dominique Eckert for comments on the manuscript and Erwin Lau for his help with the simulation data cubes used in this work. This work was supported in part by the US Department of Energy through Grant DE-FG02-91ER40681 and Purdue University. DN acknowledges support from NSF grant AST-1009811, NASA ATP grant NNX11AE07G, NASA Chandra Theory grant GO213004B, Research Corporation, and by Yale University. This work was supported in part by the facilities and staff of the Yale University Faculty of Arts and Sciences High Performance Computing Center. 


\newcommand{\noopsort}[1]{}

\end{document}